\documentclass[12pt,preprint]{aastex}
\usepackage{amssymb}
\usepackage{graphicx}

\begin{document}
\title{Observational Signatures of High-Energy Emission during the Shallow
Decay Phase of GRB X-Ray Afterglows}
\author{Y. W. Yu$^{1,3}$, X. W. Liu$^{2,4}$, and Z. G. Dai$^1$}
\affil{$^1$Department of Astronomy, Nanjing University, Nanjing
210093, China;\\ yuyw@nju.edu.cn, dzg@nju.edu.cn
\\$^2$Purple Mountain Observatory, Chinese Academy of Sciences,
Nanjing 210008, China;\\xwliu@pmo.ac.cn
\\$^3$Institute of Astrophysics, Huazhong Normal University, Wuhan
430079, China
\\$^4$National Astronomical Observatories, Chinese Academy of Sciences,
Beijing 100012, China}

\begin{abstract}
The widely existing shallow decay phase of the X-ray afterglows of
gamma-ray bursts (GRBs) is generally accepted to be due to
long-lasting energy injection. The outflows carrying the injecting
energy, based on the component that is dominative in energy, fall
into two possible types: baryon-dominated and lepton-dominated ones.
The former type of outflow could be ejecta that is ejected during
the prompt phase of a GRB and consists of a series of baryonic
shells with a distribution of Lorentz factors, and the latter type
could be an electron-positron-pair wind that is driven by the
post-burst central engine. We here provide a unified description for
the dynamics of fireballs based on these two types of energy
injection, and calculate the corresponding high-energy photon
emission by considering synchrotron radiation and inverse Compton
scattering (including synchrotron self-Compton and combined
inverse-Compton) of electrons. We find that, in the two
energy-injection models, there is a plateau (even a hump) in
high-energy light curves during the X-ray shallow decay phase. In
particular, a considerable fraction of the injecting energy in the
lepton-dominated model can be shared by the long-lasting reverse
shock since it is relativistic. Furthermore, almost all of the
energy of the reverse shock is carried by leptons, and thus the
inverse-Compton emission is enhanced dramatically. Therefore, this
model predicts more significant high-energy afterglow emission than
the baryon-dominated model. We argue that these observational
signatures would be used to discriminate between different
energy-injection models in the upcoming {\em Gamma-ray Large Area
Space Telescope} (GLAST) era.
\end{abstract}

\section{Introduction}
As discovered by Swift, there is a shallow decay phase (temporal
indices $\alpha\sim[0,-0.8]$) from post-burst several tens of
seconds to several hours (even days) occurring in the X-ray
afterglow light curves of a significant fraction of gamma-ray bursts
(GRBs) (Nousek et al. 2006; O'Brien et al. 2006; Willingale et al.
2007; Liang et al. 2007). This shallow decay phase is obviously
beyond understanding of the standard afterglow model (M\'esz\'aros
\& Rees 1997; Sari et al. 1998), but is generally accepted to be due
to continuous energy injection into relativistic blast waves (e.g.,
Zhang et al. 2006; Nousek et al. 2006; Fan \& Xu 2006; Sollerman et
al. 2007; de Pasquale et al. 2006, 2007). In particular, Liang et
al. (2007) recently compared the closure relations derived from a
simple energy-injection form, $\dot{E}_{\rm inj}\propto t^{-q}$,
with the observed temporal and spectral indices of afterglows of 53
GRBs and then argued that a roughly constant injection luminosity
may be favored by the observations. However, although this simple
injection form is usually assumed in model fitting, the specific
nature of the energy injection should be concerned and studied
further.

The physical nature of an injecting energy flow discussed in the
literature involves two possible candidates: (i) ejecta that
consists of a series of shells with a distribution of Lorentz
factors, which are dominated in energy by baryons (Rees \&
M\'esz\'aros 1998; Panaitescu et al. 1998; Sari \& M\'esz\'aros
2000; Granot \& Kumar 2006; Liu et al. 2007), and (ii) a post-burst
energy flow that results from a long-lasting activity of the central
engine (Dai \& Lu 1998a, 1998b; Zhang \& M\'esz\'aros 2001a; Wang \&
Dai 2001; Dai 2004; Fan \& Xu 2006; Yu \& Dai 2007). In the former
case, the total energy carried by the ejecta is released from the
central engine during the prompt phase of a GRB. Subsequently,
during the afterglow phase, lower-velocity shells catch up and
collide with foregoing, higher-velocity but decelerated shells
continuously because of an assumed power-law distribution of Lorentz
factors in the ejecta. This persistent collision leads to a
long-lasting Newtonian or trans-relativistic reverse shock that
propagates into the ejecta. Meanwhile, the shock-heated ejecta
pushes the outward-moving relativistic blast wave and thus reduces
the deceleration of the blast wave effectively, which accounts for
the shallow decay. In the latter case, most of the energy that
produces an afterglow is argued to be continuously released after
(not during) the burst, but the GRB ejecta may provide only a
relatively small amount of energy for the blast wave. Initially, the
post-burst energy flow may be dominated in energy by the Poynting
flux. However, as it propagates outward, the flow would evolve to an
ultrarelativistic kinetic-energy flow dominated by a component of
electron-positron pairs through some mechanisms (e.g., magnetic
reconnection) at larger radii (Coroniti 1990; Michel 1994; Kirk \&
Skj{\ae}raasen 2003). It is this ultrarelativistic leptonic wind
(rather than the electromagnetic flux) that feeds the blast wave and
maintains a long-lasting relativistic reverse shock. The emission of
the more and more energetic blast wave and the emission of the
reverse shock together give rise to the shallow decay (Dai 2004).
Therefore, we conclude from the above discussion that the injecting
energy flows are likely to be matter-dominated. Furthermore, based
on the component that is dominative in energy, the injecting flows
perhaps have two types: a baryon-dominated outflow and a
lepton-dominated outflow. The two corresponding representative
models as described above are here called the radially structured
ejecta (RSE) model (Rees \& M\'esz\'aros 1998) and the relativistic
wind bubble (RWB) model (Dai 2004), respectively.

In order to provide an effective test on the models mentioned above,
a careful investigation of high-energy emission is important and
urgent, as the launch and detection of GLAST are upcoming (Ritz
2007). In this paper, we calculate high-energy afterglow emission
during the shallow decay phase of X-ray afterglows based on these
two models and give corresponding observational signatures.
Recently, Wei \& Fan (2007), Fan et al. (2007), and Gou \&
M\'esz\'aros (2007) have made some attempts on high-energy
afterglows. In their papers, the authors studied only the effect of
the forward-shocked medium with the simple energy-injection form of
$\dot{E}_{\rm inj}\propto t^{-q}$, but didn't consider the effect of
the injecting flow itself. However, as found by Dai (2004), Uhm \&
Beloborodov (2007), Genet et al. (2007), Yu \& Dai (2007), and Liu
et al. (2007), the reverse shock that propagates into the injecting
flow could play an important role in the emission in X-ray and/or
optical bands in some situations. Therefore, high-energy emission
features due to the reverse shock are also expected. In the RSE and
RWB models, the high-energy photon emission from shocked materials
is mainly produced by inverse Compton (IC) scattering of
shock-accelerated electrons (for simplicity, electrons and positrons
are not differentiated) off synchrotron seed photons (Sari \& Esin
2001). Besides the synchrotron self-Compton (SSC) radiation from two
shocked regions, we also consider two combined IC (CIC) processes,
i.e., scattering of reverse shock photons by forward-shocked
electrons and scattering of forward shock photons by reverse-shocked
electrons, as pointed out by Wang, Dai \& Lu (2001).

The structure of this paper is organized as follows: in the next
section we provide a unified description for the dynamics of
fireballs in the RSE and RWB models. In \S3, we present the energy
distributions of shocked electrons that are determined by the
shock-acceleration effect and the synchrotron \& IC cooling effect,
and formulate calculations of the synchrotron and IC radiation
(including SSC and CIC). In \S4, we show numerical results of the
dynamics, spectra and light curves for some typical parameters in
the two models. Finally, in \S5, a summary is given and the
observability of high-energy emission by the Large Area Telescope
(LAT) instrument of GLAST is discussed.

\section{Dynamics}
As illustrated in Figure 1, in both the RWB and RSE models, the
system can be divided into four regions by the forward and reverse
shocks\footnote{It takes only a short time (tens to hundreds of
seconds) for a reverse shock to cross the ejecta of a typical GRB in
the RWB model. Meanwhile, two forward shocks forming initially
during the interaction of the ejecta both with the medium and with
the leptonic wind are assumed to merge to one forward shock. In
addition, the contribution of the GRB ejecta to the afterglow
emission should be negligible as compared with the shocked medium
and the shocked wind during the emission period of our interest.
Thus, for simplicity, the structure of an RWB can be described
approximately by Figure 1(a). However, for the RSE model, the
situation is different. The reverse shock propagates into the GRB
ejecta for a long time and thus influences the afterglow emission
during the first hours significantly (see numerical calculations in
\S 4).}: (1) the unshocked ambient medium, (2) the forward-shocked
medium, (3) the reverse-shocked materials (i.e., shocked leptonic
wind for the RWB model or shocked GRB ejecta for the RSE model), and
(4) the unshocked cold wind or GRB ejecta, where regions 2 and 3 are
separated by a contact-discontinuity surface.

\subsection{Structure of injecting flows}
Figure 1 also shows illustrative $\Gamma_i$-distributions in all the
regions. The main differences between the two models are in the
structure and composition of injecting flows:

(i) \textit{The RWB model}. Following the analysis of Dai (2004) for
a magnetar-driven wind, we simply assume that the leptonic wind
propagates outward with a constant luminosity $L_{\rm w}$ and a
constant bulk Lorentz factor $\Gamma_{\rm w}$ during a period of
$T_{\rm w}$ after the burst\footnote{A quantity $Q'$ with a prime is
measured in the comoving frame of a certain region, while the other
quantities $Q$s are measured in the local medium's rest frame except
for specific declarations. In addition, a subscript ``$i=1,2,3,4$"
represents a quantity describing region ``$i$".}. Thus, we can
calculate the number density (${n'}_4^{(\rm RWB)}$) and the bulk
Lorentz factor ($\Gamma_4^{(\rm RWB)}$) of the pre-shock materials
at the reverse shock front as
\begin{equation}
{n'}_4^{(\rm RWB)}= n'_{\rm w}(R)={L_{\rm w}\over 4\pi
R^2\Gamma_{\rm w}^2m_{\rm e}c^3},\label{n41}
\end{equation}
\begin{equation}
\Gamma_{4}^{(\rm RWB)}=\Gamma_{\rm w}, \label{g41}
\end{equation}
where $m_{\rm e}$ is the rest mass of electrons and $R$ is the
radius of the system in the thin shell approximation.

(ii) \textit{The RSE model}. As suggested by Rees \& M\'esz\'aros
(1998), the mass distribution in the wide GRB ejecta associated with
a distribution of bulk Lorentz factors reads
\begin{equation}
M_{\rm ej}(>\Gamma_{\rm ej})\propto\Gamma_{\rm ej}^{-s}.
\end{equation}
Moreover, we assume that the $\Gamma_{\rm ej}$-distribution in the
GRB ejecta satisfies
\begin{equation}
\Gamma_{\rm ej}(x)\propto x^{-1/b},
\end{equation}
where $x$ is the displacement of the reverse shock propagating into
the ejecta. If the Lorentz factor of the end of the ejecta (i.e.,
the minimum bulk Lorentz factor) is denoted by $\Gamma_{\rm
ej,min}$, then the total displacement of the reverse shock when it
crosses the ejecta could be estimated as $R_{\rm cross}/2\Gamma_{\rm
ej,min}^2\sim 10^{13}-10^{14}$ cm. Under these assumptions, the
associated injecting energy distributes with respect to $x$ as
$E_{\rm ej}\propto x^{s/b}$, and ${n'}_4^{(\rm RSE)}$ and
$\Gamma_{4}^{(\rm RSE)}$ are given by
\begin{equation}{n'}_4^{(\rm RSE)}=
n'_{\rm ej}(x)={{dM_{\rm ej}/dx}\over4\pi R^2\Gamma_{\rm
ej}(x)m_{\rm p}},\label{n42}
\end{equation}
\begin{equation}\Gamma_{4}^{(\rm RSE)}=\Gamma_{\rm
ej}(x),\label{g41}
\end{equation}
where $m_{\rm p}$ is the proton rest mass.

\subsection{Dynamic equations}
Now, we describe the dynamic evolution of the systems under the
effect of the two types of injecting flow. A resultant long-lasting
reverse shock transforms the injecting energy into the internal
energy of the reverse-shocked materials continuously. Meanwhile, a
part of fresh energy is further transformed to the kinetic energy
($E_{\rm K,2}$) of region 2 through the work done by region 3:
$\delta W=4\pi P'_3R^2dR$, where $P'_3$ is the pressure of region 3.
Then, we have
\begin{equation}
\delta W=dE_{\rm K,2}=d\left[(\Gamma_2^2-1)M_{\rm sw}c^2\right],
\end{equation}
where $M_{\rm sw}$ is the rest mass of the swept-up medium and
$\Gamma_2$ is the average bulk Lorentz factor of the shocked medium.
Thus, the dynamic evolution of region 2 is described by (Dai 2004)
\begin{equation}{d\Gamma_2\over dR}={4\pi
R^2\left[P'_3/c^2-(\Gamma_2^2-1)n_1m_{\rm
p}\right]\over2\Gamma_2M_{\rm sw}},\label{g2}
\end{equation}
where $dM_{\rm sw}/dR=4\pi R^2n_1m_{\rm p}$ is applied and $n_1$ is
the proton number density of the ambient medium. To integrate the
above equation, the pressure of region 3 should be calculated by
\begin{equation}
P'_3={1\over3}(\Gamma'_{34}-1)(4\Gamma'_{34}+3)n'_4m_{\rm
re}c^2,\label{p31}
\end{equation}where
$\Gamma'_{34}=\Gamma_{3}\Gamma_{4}(1-\beta_3\beta_4)$ is the Lorentz
factor of region 3 measured in region 4 and $\beta_i$ is the
velocity in units of $c$, and $m_{\rm re}$ represents the electron
and proton rest masses for the RWB and RSE models, respectively. The
evolution of $\Gamma_3$ can be obtained from the equality of Lorentz
factors of the two sides of the contact discontinuity surface,
assuming that the shocked medium in region 2 satisfies the adiabatic
self-similar solution of Blandford \& McKee (1976)
\begin{equation}
\Gamma_3=\Gamma_2\chi^{-1/2}.\label{g3}
\end{equation}
To fix the self-similar variable $\chi$, we use the following
relationship,
\begin{equation}P'_3={4\over3}\Gamma_2^2n_1m_{\rm p}c^2
\chi^{-17/12}.\label{p32}
\end{equation}
Combining Eqs. \ref{p31}, \ref{g3}, and \ref{p32} to eliminate
$\chi$, we can solve $\Gamma_3$ and $P'_3$ as functions of
$\Gamma_2$ and thus integrate Eq. \ref{g2}. Furthermore, the rest
mass of region 3 is obtained from
\begin{equation}
{dM_3\over dR}=4\pi R^2(\beta_4-\beta_{\rm RS})\Gamma_4n'_4m_{\rm re},
\end{equation}where
$\beta_{\rm
RS}=(\Gamma_3n'_3\beta_3-\Gamma_4n'_4\beta_4)/(\Gamma_3n'_3-\Gamma_4n'_4)$
is the velocity of the reverse shock.

When the reverse shock crosses the wind or GRB ejecta, the dynamic
evolution of region 3 is described by Kobayashi \& Sari (2000) and
Kobayashi (2000). Then, the pressure of region 3 in Eq. \ref{g2}
decreases significantly and thus is negligible. As a result, the
dynamic equation of region 2 returns to the form of the standard
afterglow model in Huang et al. (1999).

\section{Electron distributions and emission mechanisms}
After the dynamic evolution equations are given, the internal
physics of the shocked regions in the RWB and RSE models can be
considered as follows.

\subsection{Electron energy distributions}
As the forward and reverse shocks propagate, the bulk kinetic
energies of the shells are gradually transformed into the internal
energy of the shocked materials, the density of which is denoted by
$e'_i$. The internal energy will be partially carried by the
accelerated electrons and magnetic fields, the energy densities of
which are factions $\epsilon_{{\rm e},i}$ and $\epsilon_{{\rm B},i}$
of $e'_i$, respectively. Through the shock acceleration, the
electrons behind the forward/reverse shock will obtain an initial
energy distribution $N'_{{\rm ini},i}({\gamma'}_{i})\propto
{\gamma'}_{i}^{-p}$ with the minimum Lorentz factor ${\gamma'}_{{\rm
m},i}=\lambda_i[(p-2)/(p-1)]\epsilon_{{\rm
e},i}(\tilde{\Gamma}_i-1)$ and the maximum Lorentz factor
$\gamma'_{{\rm max},i}\approx10^8[B'_i(1+Y_i)]^{-1/2}$, where
$\lambda_2=m_{\rm p}/m_{\rm e}$, $\lambda_3=m_{\rm re}/m_{\rm e}$,
$\tilde{\Gamma}_2=\Gamma_{2}$, $\tilde{\Gamma}_3=\Gamma'_{34}$, and
$B'_i$ is the magnetic field strength. The occurrence of the Compton
parameter $Y_i$ is induced by the IC cooling effect. By considering
the cooling effect of the synchrotron and IC radiation, we can
obtain the
actual electron energy distributions (Huang et al. 2000):\\
for ${\gamma'}_{{\rm c},i}\leq{\gamma'}_{{\rm m},i}$,
\begin{equation}
N'_i({\gamma'}_i)\propto\left\{
\begin{array}{ll}
{\gamma'}_i^{-2}~~~~~{\rm if}~~{\gamma'}_{{\rm
c},i}\leq{\gamma'}_i\leq{\gamma'}_{{\rm m},i},\\
{\gamma'}_i^{-p-1}~~~{\rm if}~~{\gamma'}_{{\rm
m},i}<{\gamma'}_i\leq{\gamma'}_{{\rm max},i};
\end{array}\label{ng1}
\right.
\end{equation}
and for ${\gamma'}_{{\rm c},i}>{\gamma'}_{{\rm m},i}$,
\begin{equation}
N'_i({\gamma'}_i)\propto\left\{
\begin{array}{ll}
{\gamma'}_i^{-p}~~~~~{\rm if}~~{\gamma'}_{{\rm
m},i}\leq{\gamma'}_i\leq{\gamma'}_{{\rm c},i},\\
{\gamma'}_i^{-p-1}~~~{\rm if}~~{\gamma'}_{{\rm
c},i}<{\gamma'}_i\leq{\gamma'}_{{\rm max},i}.\\
\end{array}
\right.\label{ng2}
\end{equation}
The cooling Lorentz factor ${\gamma'}_{{\rm c},i}$ is defined by
equaling the cooling timescale and dynamic expansion timescale of
the system, and it reads
\begin{equation}
{\gamma'}_{{\rm c},i}={6\pi m_{\rm e}c\over(1+Y_i)\sigma_{\rm
T}\Gamma_i {B'}_i^{2}t}\label{gc}
\end{equation}
with $\sigma_{\rm T}$ being the Thomson scattering cross section.
Here, the dynamic timescale measured in the observer's frame in
calculating ${\gamma'}_{\rm c}$ is assumed to be approximately equal
to the observer's time $t$ since the trigger. In addition, the
self-absorption of the synchrotron photons is ignored in Eqs.
\ref{ng1} and \ref{ng2} for the high-energy emission that we are
interested in. Following Sari \& Esin (2001), the Compton parameter
$Y_i$ in Eq. \ref{gc}, defined as the ratio of the IC (including SSC
and CIC) luminosity to the synchrotron luminosity of electrons, is
estimated by
\begin{equation}
Y_2\equiv{L'_{\rm IC,2}\over L'_{\rm syn,2}}={U'_{\rm
syn,2}+{1\over2}U'_{\rm syn,3}\over U'_{\rm B,2}},\label{y2}
\end{equation}
\begin{equation}
Y_3\equiv{L'_{\rm IC,3}\over L'_{\rm syn,3}}={U'_{\rm
syn,3}+{1\over2}U'_{\rm syn,2}\over U'_{\rm B,3}},\label{y3}
\end{equation}
where $U'_{{\rm syn},i}=\eta_i\epsilon_{{\rm e},i}e'_{i}/(1+Y_i)$
and $U'_{{\rm B},i}=\epsilon_{{\rm B},i}e'_{i}$ are the energy
densities of synchrotron seed photons and magnetic fields,
respectively. The radiation efficiency $\eta_i$ reads
\begin{equation}
\eta_i=\left\{
\begin{array}{ll}
1,~~~~~~~~~~~~~~~~~~~~{\rm for}~~{\gamma'}_{{\rm
c},i}\leq{\gamma'}_{{\rm m},i},\\({\gamma'}_{{\rm
c},i}/{\gamma'}_{{\rm m},i})^{2-p},~~{\rm for}~~{\gamma'}_{{\rm
c},i}>{\gamma'}_{{\rm m},i}.
\end{array}\right.
\end{equation}
A factor of ${1\over2}$ in Eqs. \ref{y2} and \ref{y3} occurs,
because only about one-half of seed photons from one shocked region
will diffuse into the other one for the CIC process. In two extreme
situations, equations \ref{y2} and \ref{y3} can be simplified as \\
i) for $U'_{\rm syn,3}\gg U'_{\rm syn,2}$,
\begin{equation}
Y_2=\eta_3{\epsilon_{\rm e,3}e'_3\over2\epsilon_{\rm
B,2}e'_2}(1+Y_3)^{-1},~~Y_3=\eta_3{\epsilon_{\rm
e,3}\over\epsilon_{\rm B,3}}(1+Y_{3})^{-1};
\end{equation}
ii) for $U'_{\rm syn,3}\ll U'_{\rm syn,2}$,
\begin{equation}
Y_2=\eta_2{\epsilon_{\rm e,2}\over\epsilon_{\rm
B,2}}(1+Y_{2})^{-1},~~Y_3=\eta_2{\epsilon_{\rm
e,2}e'_2\over2\epsilon_{\rm B,3}e'_3}(1+Y_2)^{-1}.
\end{equation}

\subsection{SSC and CIC emission}
Once the electron distribution is known, the synchrotron emissivity
of electrons in region $i$ at frequency $\nu'$ is calculated
directly by (Rybicki \& Lightman 1979)
\begin{equation}
{\varepsilon'}_i^{\rm syn}(\nu')={\sqrt{3}q_{\rm e}^3B'_i\over
m_{\rm e}c^2}\int d{\gamma'}_iN'_i({\gamma'}_i)\mathcal
{F}\left({\nu'\over\nu'_{\rm c}}\right),\label{psyn}
\end{equation}
where $q_{\rm e}$ is the electron charge, $\nu'_{\rm
c}=3{\gamma'}_i^2q_{\rm e}B'_i/(4\pi m_{\rm e}c)$,
$\mathcal{F}(u)=u\int_u^{\infty}K_{5/3}(k)dk$, and $K_{5/3}(k)$ is
the Bessel function.

Accompanying with the synchrotron radiation, the electrons also lose
their energy through upscattering the synchrotron seed photons. As
usual, the first-order IC scattering is considered and the
higher-order processes are neglected. According to Blumenthal \&
Gould (1970), when the Klein-Nishina suppression is considered, the
IC emissivity (at a frequency $\nu'$) of electrons in region $i$
upscattering seed photons from region $j$ is calculated by ($i=j$
for SSC and $i\ne j$ for CIC)
\begin{equation}
{\varepsilon'}_i^{{\rm IC}(j)}(\nu')=3\sigma_{\rm T}\int
d{\gamma'}_iN'_i({\gamma'}_i)\int
{d{\nu}_{j}^{\dag}}{\nu'f_{j}^{\dag}\over
4{\gamma'}_i^2{{\nu}_{j}^{\dag}}^2}g(x,y),
\end{equation}
where $g(x,y)=2y\ln y+(1+2y)(1-y)+{1\over2}{x^2y^2\over1+xy}(1-y)$
with $x=4\gamma'_ih{\nu}_{j}^{\dag}/m_{\rm e}c^2$,
$y={h\nu'/[x(\gamma'_im_{\rm e}c^2-h\nu')]}$. ${\nu}_{j}^{\dag}$ and
${f}_{j}^{\dag}$ are the frequency and the corresponding flux
density of the incident photons from region $j$, respectively, which
are measured in the comoving frame of region $i$.

The observed synchrotron and IC flux densities at a frequency $\nu$
(measured in the observer's frame) from region $i$ are given
respectively by (Huang et al. 2000)
\begin{equation}
F_{\nu,i}^{\rm syn}(t)=\int_0^{\pi}d\theta V'_i{\sin
\theta\over2}{{\varepsilon'}_i^{\rm syn}(\mathcal{D}_i\nu)\over 4
\pi D_{\rm L}^2\mathcal{D}_i^3},\label{fsyn}
\end{equation}
\begin{equation}
F_{\nu,i}^{{\rm IC}(j)}(t)=\int_0^{\pi} d\theta
V'_i{\sin\theta\over2}{{\varepsilon'}_i^{{\rm
IC}(j)}(\mathcal{D}_i\nu)\over4 \pi D_{\rm
L}^2\mathcal{D}_i^3},\label{fic}
\end{equation}
where $D_{\rm L}$ is the luminosity distance and
$\mathcal{D}_i=\Gamma_i(1-\beta_i\cos\theta)$ is the Doppler factor.
Here, the viewing angle is assumed to be zero and $\theta$ is the
angle between the moving direction of the emitting material and the
line of sight. The volumes $V'_i$ of the shocked regions in Eqs.
\ref{fsyn} and \ref{fic} are functions of $\theta$ because the
integrations are performed over the equal-arrival surface. In
addition, the jet correction should be considered if the GRB ejecta
is collimated.

\section{Numerical Results}
\subsection{Dynamic evolution}
For the RWB and RSE models, we first calculate the dynamic evolution
with radius. The initial value of $R$ is taken to be $\sim10^{16}$
cm as the deceleration radius. For the RWB model, as in Yu \& Dai
(2007), the isotropic wind luminosity is $L_{\rm
w}=4\times10^{47}B_{14}^2~\rm erg~s^{-1}$, the wind duration $T_{\rm
w}=5\times10^4B_{14}^{-2}$ s, and the bulk Lorentz factor of the
wind $\Gamma_{\rm w}=10^4$, where $B_{14}$ is the magnetic field
strengthen of the central magnetar in unit of $10^{14}$ G. Here, we
take $B_{14}=4$ for a typical magnetar. Thus the total energy of the
wind is $E_{\rm w}=2.0\times10^{52}$ erg. However, the isotropic
kinetic energy of the previous GRB ejecta is relatively small,
$E^{\rm RWB}_{\rm ej}=10^{51}$ erg. For consistency, we assume that,
in the RSE model, the isotropic kinetic energy carried by the wide
GRB ejecta is $E^{\rm RSE}_{\rm ej}=2.1\times10^{52}$ erg, and the
distribution-related parameters are $\Gamma_{\rm ej,min}=30$,
$\Gamma_{\rm ej, max}=500$, and $s=b=1.5$. In addition, the number
density of the ambient interstellar medium is $n_1=1~\rm cm^{-3}$ in
both cases.

The dotted lines in Figure 2 represent the dynamic evolution of
blast waves with a certain energy (i.e., without energy injection),
and the solid lines show the energy injection case. We can see
clearly from this figure that the energy ($\propto\Gamma_2^2$) of
the shocked medium increases gradually until the energy injection is
over. As analyzed in Yu \& Dai (2007), the final energy carried by
the shocked medium is a fraction $\sim 67\%$ of the total injecting
energy for the RWB model versus $\sim 90\%$ for the RSE model.
Meanwhile, the other fraction of the injecting energy ($\sim 33\%$
and $\sim 10\%$ for the RWB and RSE models, respectively) should be
shared by the reverse-shocked material. In addition, we can know
that the reverse shock is relativistic in the RWB model, but
Newtonian or trans-relativistic in the RSE model, according to an
estimation of the Lorentz factor of region 3 relative to region 4
from
\begin{equation}
\Gamma'_{34}\approx{1\over2}\left({\Gamma_3\over\Gamma_4}
+{\Gamma_4\over\Gamma_3}\right)
\end{equation}
with the values of $\Gamma_3$ as shown in Figure 2. This difference
between the reverse shocks in the two models is just the reason why
the reverse shock in the RWB model can share more injecting energy
than the one in the RSE model.

\subsection{Spectra and light curves}
To calculate the emission of shocked materials, we take the
microphysical parameters $p=2.3$ and $\epsilon_{\rm B}=0.01$ for all
shocked regions. Then, $\epsilon_{\rm e}$ is calculated by
$\sqrt{\epsilon_{\rm B}}$ for the baryon-dominated regions (i.e.,
regions 2 \& 3 in the RSE model and region 2 in the RWB model) as
argued by Medvedev (2006) and $(1-\epsilon_{\rm B})$ for the
lepton-dominated region (i.e., region 3 in the RWB model). The
luminosity distance of a GRB is taken to be $D_{\rm L}=1$ Gpc,
corresponding to a redshift of 0.2.

Figure 3 shows the light curves of an RWB. From this figure, we
obtain the following results. First, the shallow decay phase of an
X-ray (keV) afterglow is produced during the early hours (left upper
panel), which is mildly dominated by the reverse shock emission (for
a detailed discussion see Yu \& Dai 2007). The synchrotron radiation
is the dominative mechanism in this band. Second, an obvious hump
accompanying with the X-ray plateau occurs in a high-energy (MeV \&
GeV) afterglow, which is dominated by the IC (especially SSC)
emission from the reverse-shocked wind. Finally, a comparison of
light curves in different bands (right lower panel) shows that the
high-energy (especially GeV) emission flux is mildly or even
significantly higher than the one in X-ray band. This feature is
also implied obviously by the spectra as shown in Figure 4. The peak
flux of the IC emission at $\sim1$ GeV is about one order of
magnitude higher than the synchrotron peak at $\sim10$ eV. Moreover,
the IC component is dominative above sub-MeV and, in all bands, the
emission from the reverse-shocked wind is more important than the
one from the forward-shocked medium at 1000 s.

The results for the RSE model are shown in Figures 5 and 6. From
these two figures, we find that, first, the X-ray shallow decay
phase during the early hours is also produced in this model as shown
in left upper panel of Figure 5, which is dominated by the forward
shock. Second, the high-energy afterglow light curves also have a
flattening segment. However, differing from the RWB model, the
contribution to the high-energy plateau from the IC emission plays a
role only in GeV band, whereas the MeV emission is totally
contributed by the synchrotron mechanism during the total afterglow
phase. This result arises from the relative weakness of the IC
component in the RSE model as shown in Figure 6. Therefore, the flux
in MeV or GeV band is lower than or approximately equal to the X-ray
flux as shown in right lower panel of Figure 5. In addition, Figure
6 shows that the contribution of the emission from the
reverse-shocked ejecta is negligible in all bands for the parameters
that we adopt.

In conclusion, we obtain very different high-energy afterglows in
the RWB and RSE models with a same amount of injecting energy that
gives rise to similar X-ray afterglows. The RWB model predicts more
significant (about one order of magnitude stronger) high-energy
emission than the RSE model. The reasons for this difference are
that the reverse shock in the RWB model is relativistic and that the
energy of the reverse shock is almost totally carried by electrons.
The former reason enables the reverse shock to share more (i.e.,
33\% vs 10\%) injecting energy as discussed in Sect. 4.1, and the
latter reason dramatically increases the emission efficiency of the
reverse-shocked material through the enhanced IC emission.
Therefore, a higher fraction of injecting energy can be radiated
from the shocked regions in the RWB model. In the baryon-dominated
injection model, however, this significant enhanced IC component
doesn't happen even if the emission flux from the reverse shock
exceeds the one from the forward shock under some extreme
conditions, because most of the injecting energy in both shocked
regions is locked in the baryons whose emission is weak. Therefore,
we argue that the difference in high-energy emission between the RWB
and RSE models is essentially due to the physical distinction
between the two types of energy injection. Thus, reasonable
variations of the parameters for the models should not change our
results significantly.

\section{Summary and discussion}
The discovered shallow decay phase of GRB X-ray afterglows suggests
long-lasting energy injection into relativistic blast waves. The
injecting flow is likely to be dominated in energy by the component
of either baryons or leptons. In this paper, we first provide a
unified description for dynamics and radiation in two representative
models, i.e., the RSE and RWB models. Through maintaining a
long-lasting reverse shock and doing work to forward-shocked medium,
both types of energy injection can produce the flattening segment in
X-ray afterglow light curves easily. Second, we pay attention to
calculations of the simultaneous high-energy emission that is due to
synchrotron and, especially, IC (including SSC and CIC processes)
radiation from the shocked materials. Our results show that, during
the shallow decay phase of X-ray afterglows, there is a plateau
(even a hump) in high-energy light curves in both the RSE and RWB
models. As argued by Wei \& Fan (2007), we suggest that the
plateau/hump might account for the delayed high-energy emission of
some bursts such as GRB 940217.

We also find that the high-energy emission derived from the two
models has different observable features, e.g., morphologies of the
light curves and spectra. In particular, more significant
high-energy emission is predicted by the RWB model, because more
injecting energy in this model is carried by electrons and thus the
IC scattering is enhanced significantly. This difference in
high-energy emission between the two models could be tested in the
upcoming GLAST era. As proposed by Zhang \& M\'esz\'aros (2001b) and
Gou \& M\'esz\'aros (2007), the fluence threshold of the LAT
instrument aboard GLAST for an effective observation time $t_{\rm
eff}$ in an observation energy band centered around a photon energy
$E$ is estimated by $F_{\rm thr}\sim [\Phi_0(T/t_{\rm
eff})^{1/2}]Et_{\rm eff}$, where the flux sensitivity,
$[\Phi_0(T/t_{\rm eff})^{1/2}]$, scales as $t_{\rm eff}^{-1/2}$ due
to a limitation by the background for a long-time observation and
$\Phi_0$ is the sensitivity over an exposure time $T$. Specifically,
over a one-year exposure period, the sensitivity above 100 MeV is
$\sim 1.33\times10^{-9}~\rm ph~s^{-1}~cm^{-2}$ for point-source
observations of GRB afterglows (Gou \& M\'esz\'aros 2007). On the
other hand, for a short-time observation, the fluence threshold is
instead calculated by $F_{\rm thr}=5E/A_{\rm eff}$ with the
assumption that at least 5 photons are collected on an effective
area $A_{\rm eff}$ of the instrument. Taking $E=400$ MeV and $A_{\rm
eff}=6000~ \rm cm^{-2}$, we obtain the fluence threshold of GLAST
LAT,
\begin{equation}
F_{\rm thr}=\left\{
\begin{array}{ll}
5.3\times10^{-7}~~~~~{\rm erg~cm^{-2}},~~~t\leq2.43\times 10^4\,{\rm s},\\
3.4\times10^{-9}t^{1/2}~{\rm erg~cm^{-2}},~~~t>2.43\times 10^4\,{\rm s},
\end{array}\right.
\end{equation}
where an observation efficiency $\eta\equiv t_{\rm eff}/t=0.5$ due
to the occultation by the earth is considered. With this fluence
threshold, the observability of high-energy emission in the RWB and
RSE models by GLAST LAT is shown in Figure 7. It can be seen that
both the high-energy emissions predicted by the two models can be
detected by GLAST LAT for the given parameters. The partial fluence
of the high-energy emission in the RWB model is about an order of
magnitude higher than that in the RSE model for a same amount of
injecting energy. Here, the partial fluence is defined as an
integration of the flux density ($F_{\nu}$) over the GLAST LAT
energy band [20 MeV, 300 GeV] and the time intervals [$0.5t, t$] as
done in Gou \& M\'esz\'aros (2007).

Although the detection of the RWB model by GLAST LAT is obviously
easier, joint observations of GLAST and Swift are necessary to
distinguish between the RWB and RSE models. Simultaneous
observations of a high-energy afterglow by GLAST LAT and an X-ray
afterglow by Swift XRT for a same GRB could provide a ratio of
fluxes ($\nu F_{\nu}$) in the high-energy band to the X-ray. As
implied by the right lower panels of Figures 3 and 5, for example,
the injecting flow could be dominated in energy by baryons if the
ratio $(\nu F_{\nu})_{\rm GeV}/(\nu F_{\nu})_{\rm keV}$ is close to
1. On the other hand, if the ratio is significantly higher than 1,
there could be lepton-dominated energy injection.

\section*{Acknowledgements}
We would like to thank Lijun Gou and an anonymous referee for
valuable comments and suggestions that have helped us to improve an
earlier version of this manuscript. This work is supported by the
National Natural Science Foundation of China (grant no. 10221001 and
10640420144). YWY is also supported by the Visiting PhD Candidate
Foundation of Nanjing University and the National Natural Science
Foundation of China (grant no.10603002).

%
\begin{figure}\resizebox{\hsize}{!}{\includegraphics{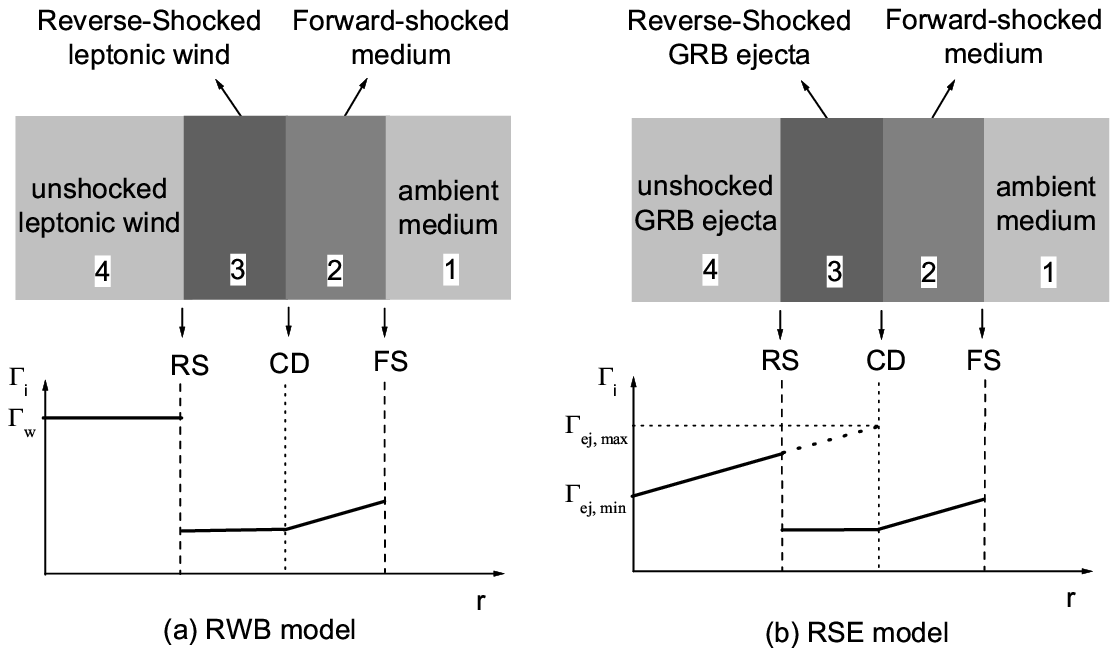}}
\caption{Illustrations of four regions divided by the forward and
reverse shocks in the RWB (left) and RSE (right) models, and the
corresponding distributions of Lorentz factors (not scaled).}
\end{figure}
%
%
\begin{figure}\resizebox{\hsize}{!} {\includegraphics{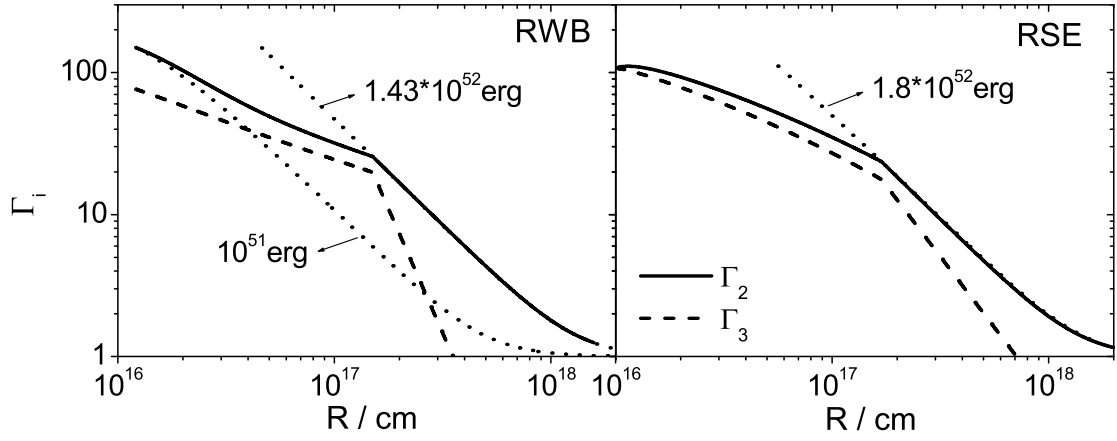}}
\caption{Evolution of the Lorentz factors of regions 2 and 3 as the
shells in the RWB (left) and RSE (right) models expand. A dotted
line represents the dynamic evolution of the shell with a fixed
amount of energy as labeled.}
\end{figure}
%
%
\begin{figure}
\resizebox{\hsize}{!} {\includegraphics{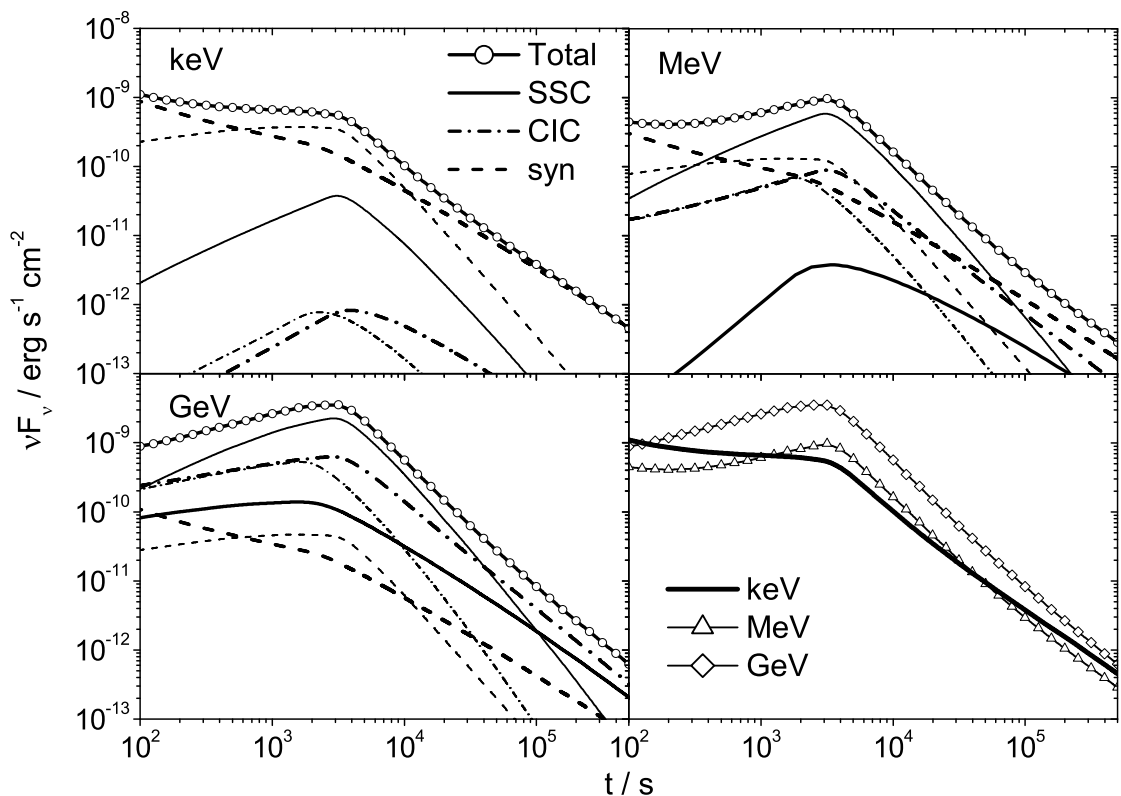}} \caption{Light
curves in X-ray (keV) and high energy gamma-ray (MeV, GeV) bands for
the RWB model. The emission from the forward-shocked medium is
represented by thick lines, while the reverse shock corresponds to
thin lines. The solid, dash-dotted, and dashed lines are due to SSC,
CIC, and synchrotron processes, respectively.}
\end{figure}
%
%
\begin{figure}
\resizebox{\hsize}{!}{\includegraphics{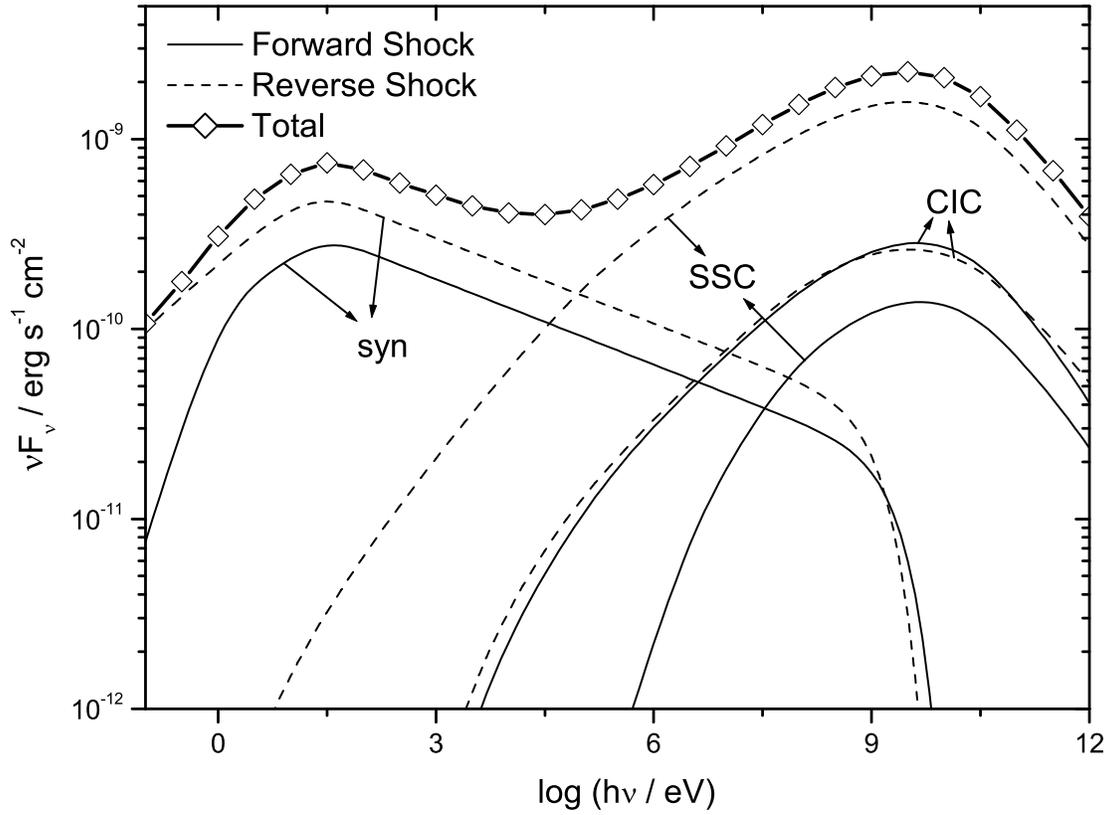}}
\caption{Synchrotron and IC spectra at 1000 s for the RWB model. The
solid lines represent the emission contributed by the
forward-shocked medium, while the dotted lines correspond to the
reverse-shocked wind.}
\end{figure}
%
%
\begin{figure}
\resizebox{\hsize}{!} {\includegraphics{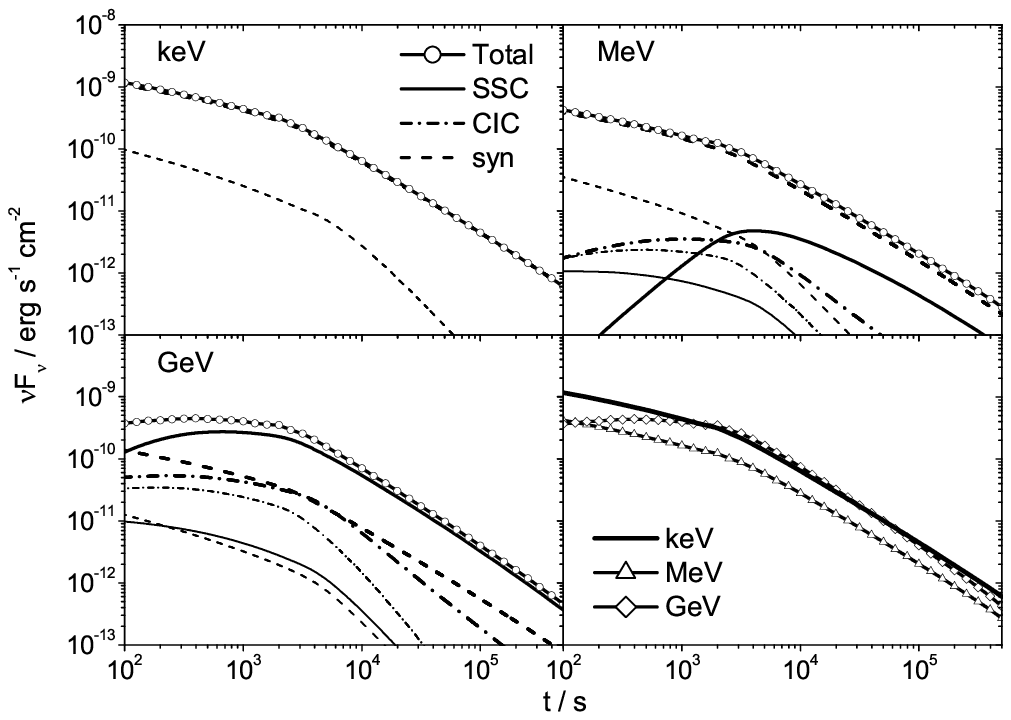}} \caption{Same as
in Figure 3 but for the RSE model.}
\end{figure}
%
%
\begin{figure}
\resizebox{\hsize}{!} {\includegraphics{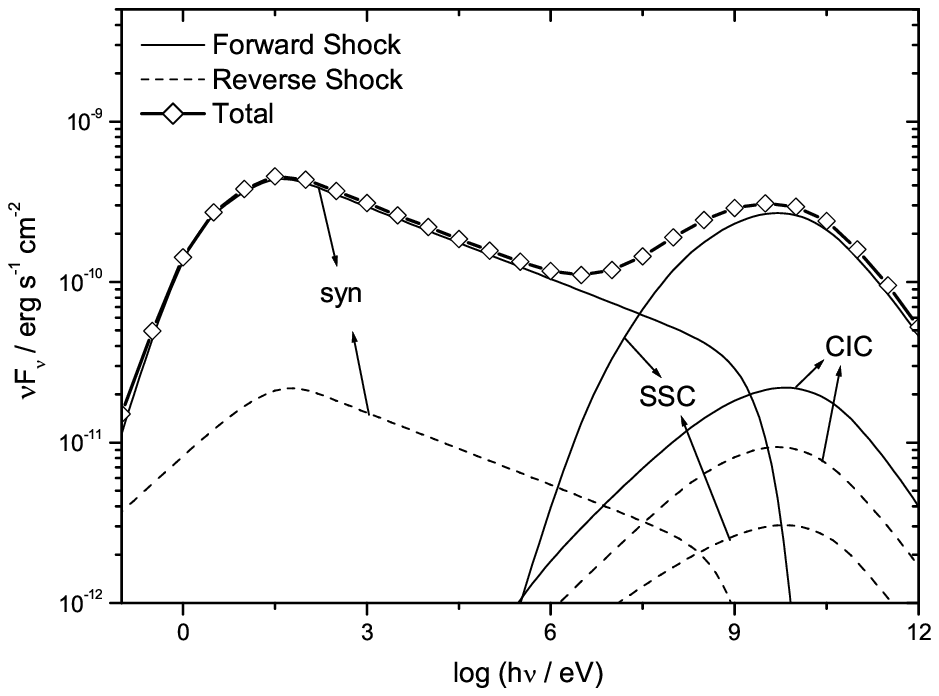}} \caption{Same as
in Figure 4 but for the RSE model.}
\end{figure}
%
%
\begin{figure}
\resizebox{\hsize}{!} {\includegraphics{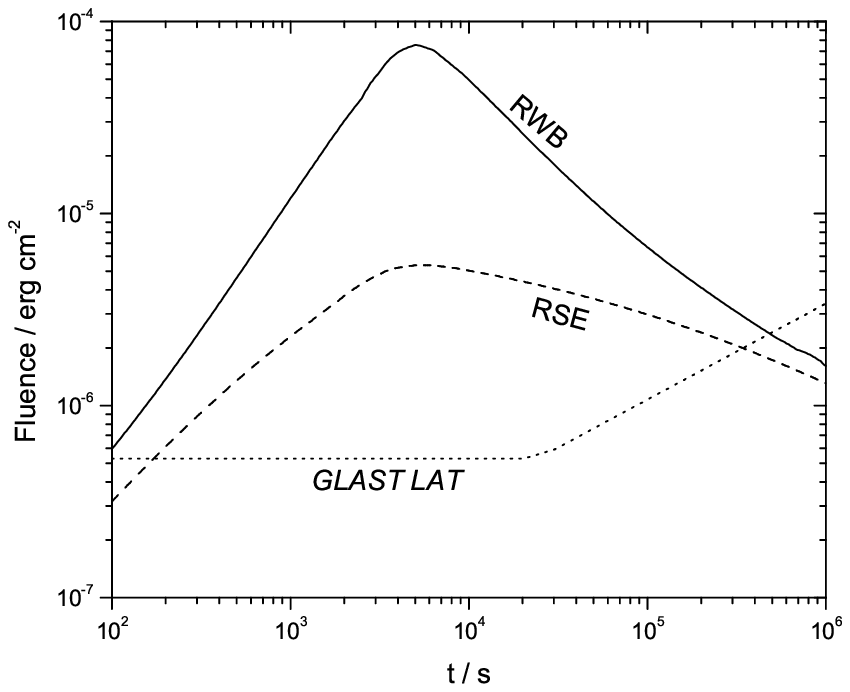}} \caption{The
partial fluence (defined as
$\int_{0.5t}^{t}\int_{\nu_1}^{\nu_2}F_{\nu}d\nu dt$, where $h\nu_1$
= 20 MeV and $h\nu_2$ = 300 GeV) curves for the RWB and RSE models.
The dotted line represents the fluence threshold of the GLAST LAT at
400 MeV.}
\end{figure}
%

\begin{thebibliography}{}
\bibitem{}Blandford, R. D., \& McKee, C. F. 1976, Phys. Fluids., 19, 1130
\bibitem{}Blumenthal, G. R., \& Gould, R. J. 1970, Rev. Mod. Phys., 42, 237
\bibitem{}Coroniti, F. V. 1990, ApJ, 349, 538
\bibitem{}Dai, Z. G. 2004, ApJ, 606, 1000
\bibitem{}Dai, Z. G., \& Lu, T. 1998a, A\&A, 333, L87
\bibitem{}Dai, Z. G., \& Lu, T. 1998b, Phys. Rev. Lett., 81, 4301
\bibitem{}de Pasquale, M., et al. 2006, MNRAS, 365, 1031
\bibitem{}de Pasquale, M., et al. 2007, accepted by MNRAS, arXiv:astro-ph/0703447
\bibitem{}Fan, Y. Z., \& Xu, D. 2006, MNRAS, 372, L19
\bibitem{}Fan, Y. Z., Piran, T., Narayan, R., \& Wei, D. M., 2007, submitted to MNRAS, arXiv:0704.2063
\bibitem{}Genet, F., Daigne, F., \& Mochkovitch, R. 2007, arXiv:astro-ph/0701204
\bibitem{}Gou, L. J., \& $\rm M\acute{e}sz\acute{a}ros$, P. 2007, accepted by ApJ, arXiv:0705.1545
\bibitem{}Granot, J., \& Kumar, P., 2006, MNRAS, 366, L13
\bibitem{}Huang, Y. F., Dai, Z. G., \& Lu, T. 1999, MNRAS, 309, 513
\bibitem{}Huang, Y. F., Gou, L. J., Dai, Z. G. \& Lu, T. 2000, ApJ, 543, 90
\bibitem{}Kirk, J. G., \& Skj{\ae}raasen, O. 2003, ApJ, 591, 366
\bibitem{}Kobayashi, S. 2000, ApJ, 545, 807
\bibitem{}Kobayashi, S., \& Sari, R. 2000, ApJ, 542, 819
\bibitem{}Liang, E. W., Zhang, B. B., \& Zhang, B. 2007, accepted by ApJ, arXiv:0705.1373
\bibitem{}Liu, X. W., Wu, W. F., Zou, Y. C., \& Lu, T. 2007, to be submitted
\bibitem{}Medvedev, M. V. 2006, ApJ, 651, L9
\bibitem{}$\rm M\acute{e}sz\acute{a}ros$, P., \& Rees, M. J. 1997, ApJ, 476, 232
\bibitem{}Michel, F. C. 1994, ApJ, 431, 397
\bibitem{}Nousek, J. A., et al. 2006, ApJ, 642, 389
\bibitem{}O'Brien, P. T., et al. 2006, ApJ, 647, 1213
\bibitem{}Panaitescu, A., $\rm M\acute{e}sz\acute{a}ros$, P., \& Rees, M. J. 1998, ApJ, 503, 314
\bibitem{}Rees, M. J., \& $\rm M\acute{e}sz\acute{a}ros$, P. 1998, ApJ, 496, L1
\bibitem{}Ritz, S. 2007, a review talk in the First GLAST Symposium (5 February 2007)
\bibitem{}Rybicki, G. B., \& Lightman, A. P. 1979, Radiative Processes in Astrophysics (New York: Wiley)
\bibitem{}Sari, R., \& Esin, A. A. 2001, ApJ, 548, 787
\bibitem{}Sari, R., Piran, T., \& Narayan, R. 1998, ApJ, 497, L17
\bibitem{}Sari, R., \& $\rm M\acute{e}sz\acute{a}ros$, P. 2000, ApJ, 535, L33
\bibitem{}Sollerman, J., et al. 2007, A\&A, 466, 839
\bibitem{}Uhm, Z. L., \& Beloborodov, A. M. 2007, ApJ, 665, L93
\bibitem{}Wang, W., \& Dai, Z. G. 2001, Chin. Phys. Lett., 18, 1153
\bibitem{}Wang, X. Y., Dai, Z. G., \& Lu, T., 2001, ApJ, 556, 1010
\bibitem{}Wei, D. M., \& Fan, Y. Z. 2007, ChJAA, 7, 509
\bibitem{}Willingale, R., et al. 2007, ApJ, 662, 1093
\bibitem{}Yu, Y. W., \& Dai, Z. G. 2007, A\&A, 470, 119
\bibitem{}Zhang, B., et al. 2006, ApJ, 642, 354
\bibitem{}Zhang, B., \& $\rm M\acute{e}sz\acute{a}ros$, P. 2001a, ApJ, 552, L35
\bibitem{}Zhang, B., \& $\rm M\acute{e}sz\acute{a}ros$, P. 2001b, ApJ, 559, 110
\end{thebibliography}
\end{document}